# Terahertz-Mediated Microwave-to-Optical Transduction

Furkan Sahbaz[1,3], James N. Eckstein[2,3], Dale J. Van Harlingen[2,3] and Simeon I. Bogdanov[1,3]

[1]*Department of Electrical and Computer Engineering, and Nick Holonyak, Jr. Micro and Nanotechnology Laboratory, University of Illinois at Urbana-Champaign, Urbana, Illinois 60801, USA*

[2]*Department of Physics and Materials Research Laboratory, University of Illinois at Urbana Champaign, Urbana, Ilinois 81801*

[3]*Illinois Quantum Information Science and Technology Center, University of Illinois Urbana-Champaign, Urbana, Illinois 61801, USA*

Transduction of quantum signals between the microwave and the optical ranges will unlock powerful hybrid quantum systems enabling information processing with superconducting qubits and low-noise quantum networking through optical photons. Most microwave-to-optical quantum transducers suffer from thermal noise due to pump absorption. We analyze the coupled thermal and wave dynamics in electro-optic transducers that use a two-step scheme based on an intermediate frequency state in the THz range. Our analysis, supported by numerical simulations, shows that the two-step scheme operating with a continuous pump offers near-unity external efficiency with a multi-order noise suppression compared to direct transduction. As a result, two-step electro-optic transducers may enable quantum noise-limited interfacing of superconducting quantum processors with optical channels at MHz-scale bitrates.

Superconducting qubits are prime candidates for scalable quantum information processing, while optical photons offer long-distance transmission of quantum states in ambient conditions. Faithful conversion of quantum states between microwave and optical frequency ranges, called quantum transduction, is instrumental in bridging the gap between these complementary functionalities. For example, optical interconnects transmitting information between individual superconducting quantum processors will enable distributed quantum computing. Conversely, superconducting quantum processors integrated into quantum optical networks will equip the latter with deterministic quantum gates and error correction [1].

The key figures of merit for a microwave-to-optical quantum transducer are the conversion efficiency $\eta$, quantum state fidelity F, and bandwidth $\Delta\omega$ [2-4]. $\eta$ and F are dictated by the error-correction thresholds, while $\Delta\omega$ must accommodate the qubit radiative rates, typically in the MHz range [1,5]. In recent experiments, microwave-to-optical quantum transduction has been realized with optomechanical [6-9], spin wave [10-12], atom-assisted [13-15] and electro-optic (EO) [16-24] interfaces. However, no platform to date has reached quantum-limited microwave-to-optical transduction, i.e. combine a near-unity efficiency with added noise below the single photon level ($n_\mu \ll 1$) at MHz bandwidths.

EO microwave-to-optical transducers stand out due to their bandwidths of up to tens of MHz and integration-friendly architectures. They are realized by combining nonlinear optical materials with 3D [18-20] or planar [21-24] superconducting microwave resonators. A strong optical pump $p_o$ propagating in the nonlinear material mediates the interaction between a microwave mode $s_\mu$ and an optical sideband mode $s_o$ as in Fig. 1(a). Due to weak nonlinearities in the optical range and the large wavelength mismatch, pump powers exceeding 1 W are typically required for efficient EO conversion. Even the absorption of a minute fraction of pump photons leads to a prohibitively strong thermal microwave photon generation and excess losses in optical and microwave modes, affecting both $\eta$ and $n_\mu$. Recent works on mm-wave superconducting circuits suggest that they can be a viable stepping stone for microwave-to-optical transduction [15,25,26]. Most recently, mm-wave-to-optical transduction using neutral atoms has shown attractive noise and efficiency figures of merit [15]. In this letter, we show that transducing photons from the microwave to the optical range through an intermediate high-frequency state can help overcome the fundamental limitations of single-step EO transducers. Specifically, we analyze a scenario where the microwave photon is first upconverted to an intermediate frequency by using kinetic inductance nonlinearity in niobium nitride (NbN), then converted through the second-order optical nonlinearity in lithium niobate (LN). By studying the noise scaling as a function of the intermediate frequency $\omega_i$ and waveguide dimensions, we show that this two-step transduction reduces the thermal noise by orders of magnitude while maintaining near-unity conversion efficiency. We confirm the model validity by numerically simulating the transduction from 8 GHz to 200 THz through an intermediate frequency state in a NbN-based two-step converter. The proposed approach promises quantum transduction with quantum-limited performance at MHz-scale conversion rates.

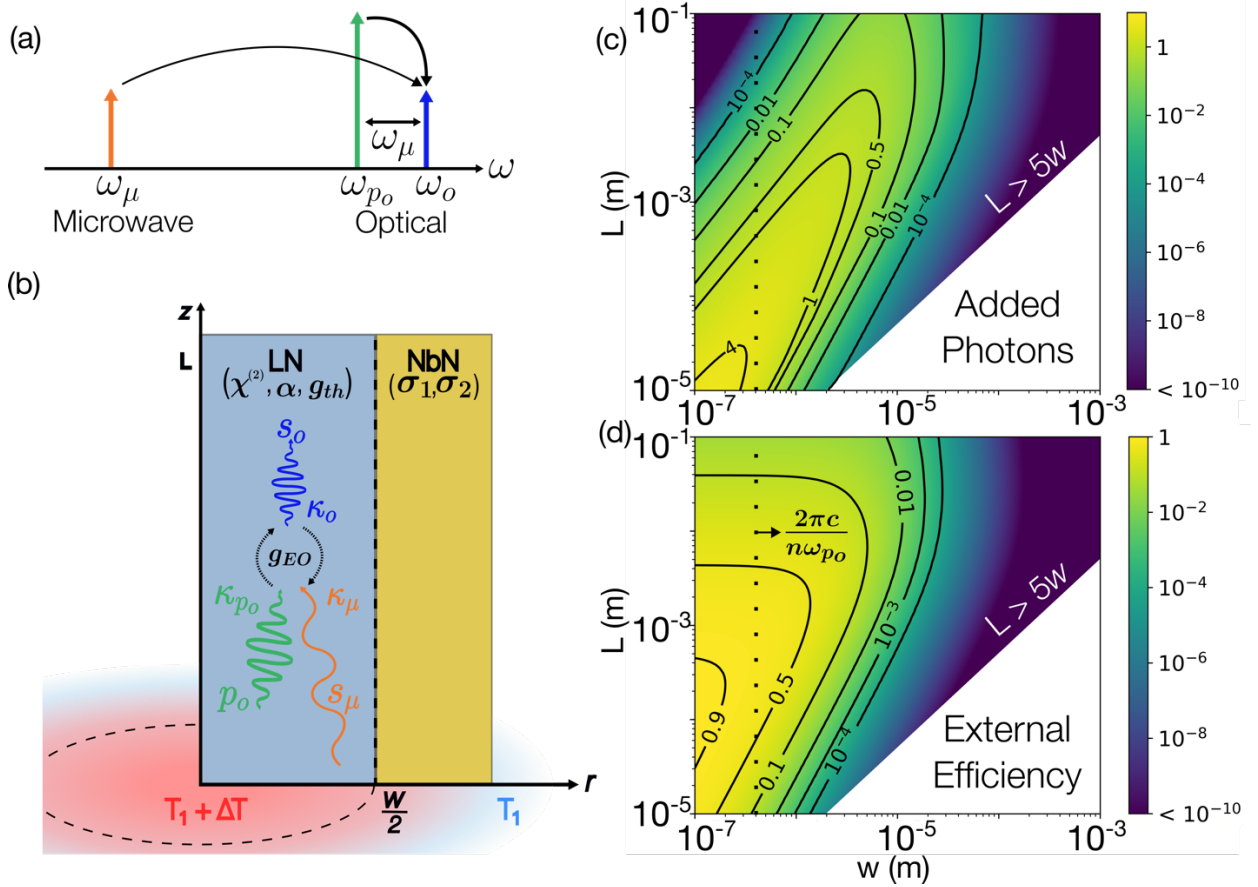

FIG. 1. (a) Single-step quantum transduction in the frequency domain. (b) Schematics of a single-step microwave-to-optical transducer model consisting of a LN optical waveguide of cross-section diameter $w$ and length $L$ integrated with a NbN microwave waveguide. The heating $\Delta T$ is caused by optical pump absorption in LN. The thermal transport is assumed to be radial. (c) The microwave mode occupancy and (d) the external conversion efficiency as a function of $w$ and $L$ for $\omega_\mu = 2\pi \times 8\ GHz$ and base temperature $T_1 = 10$ mK. The dashed lines represent the optical wavelength, where $n$ and $c$ are the LN refractive index at $\omega_{p_o}$ and the speed of light, respectively.

*Heat scaling in electro-optic transducers.* First, we analyze the noise and efficiency scaling laws of single-step EO conversion mediated by electro-optic nonlinearity ($\chi^{(2)}$). Specifically, we calculate the thermal microwave mode occupancy, $n_{\mu,1}$, and the external transduction efficiency, $\eta_1$ in a travelling-wave converter according to a simple analytical model. We consider a microwave NbN waveguide of length $L$ and width $w$, with an embedded optical LN waveguide of width $w$, where the pump signal $p_o$ (at $\omega_{p_o} = 2\pi \times 200$ THz) mixes with $s_\mu$ (at $\omega_\mu = 2\pi \times 8$ GHz) to obtain the sideband $s_o$ as in Fig. 1(a) and 1(b). We assume unity electro-optic cooperativity ($C_{EO} = 4g_{EO}^2|p_o|^2/(\kappa_\mu \kappa_o) = 1$), where $g_{EO}, |p_o|^2, \kappa_\mu$ are $\kappa_o$ are the EO coupling rate, number of pump photons, and loss rates for microwave and optical modes (see Supplemental Material, Chapter I). The total loss rates of all the modes consist of internal (absorption) and external (outcoupling) loss rates: $\kappa_\mu = \kappa_{\mu,i} + \kappa_{\mu,e}$, $\kappa_o = \kappa_{o,i} + \kappa_{o,e}$ and $\kappa_o = \kappa_{p,i} + \kappa_{p,e}$ [17]. We also assume unity spatial overlap and perfect phase matching between the modes $p_o$ and $s_\mu$ and $s_o$. Explicit values and expressions of all the geometric and material parameters involved are given in Supplemental Material, Chapter II.

Under the boundary condition $C_{EO} = 1$, the total efficiency becomes equal to the external efficiency $\eta_1 = \kappa_{\mu,e} \kappa_{o,e}/(\kappa_\mu \kappa_o)$. The thermal microwave mode occupancy is calculated by assuming the microwave mode is coupled to the cryostat environment with base temperature $T_1$ and the bath of the active medium (LN) at a temperature $T_1 + \Delta T$ heated by pump absorption. The thermal Bose-Einstein occupation factors due to the cryostat environment ($n_F(T_1)$) and the heated active medium ($n_B(T_1 + \Delta T)$) add up to the total thermal microwave mode occupancy $n_{\mu 1} = n_B(T_1 + \Delta T) \kappa_{\mu,e}/\kappa_\mu + n_F(T_1) \kappa_{\mu,i}(T_1 + \Delta T)/\kappa_\mu$ [27,28].

$n_{\mu 1}$ and $\eta_1$ are thus determined by active medium temperature $T_1 + \Delta T$ through the strongly temperature-dependent microwave propagation losses and the Bose-Einstein occupation factor. The heating $\Delta T$ is induced by the uniformly distributed pump photons with total energy $|p_o|^2 \hbar \omega_{p_o}$ being absorbed at a rate $\kappa_{p,i}$. We assume radial heat transfer from the transducer's nonlinear waveguide to the cladding material, which is a good approximation when $L \gg w$. In this model, we describe the heat transport with a common thermal conductivity value $g_{th}(T_1 + \Delta T)$ equal to that of LN. We calculate the heating in the regime of continuous wave (CW) pumping guaranteeing maximum transduction bandwidth, which leads to $\Delta T = |p_o|^2 \hbar \omega_{p_o} \kappa_{p,i}/g_{th}L$. Under the $C_{EO} = 1$ constraint, $\Delta T$ obeys the following steady-state equation [29,30]:

$$\Delta T = \frac{\kappa_\mu(T_1 + \Delta T)\kappa_o}{4 g_{EO}^2} \cdot \frac{\hbar \omega_{p_o} \kappa_{p_o,i}}{g_{th}(T_1 + \Delta T)L} \quad (1)$$

This implicit equation highlights the difficulty of achieving quantum-limited performance. Heating leads to excess microwave loss $\kappa_\mu$, which, through the "$C_{EO} = 1$" constraint, requires a higher pump power, leading to even more heating. Low operating temperatures present an additional challenge as the thermal conductivities of most materials vanish (see Supplemental Materia, Chapter III). Figures 1(c) and 1(d) plot $\eta_1$ and $n_{\mu,1}$ as a function of transducer dimensions $w$ and $L$, at $T_1 = 10$ mK. In agreement with previous works [18-24], increasing $L$ improves the efficiency due to reduced heating and correspondingly lower absorption. Eventually, as $L$ tends to infinity, the propagation losses take over and the efficiency reduces. Narrowing $w$ improves the coupling rate $g_{EO}$, reducing the optical pumping requirements, and exponentially suppressing the noise photon population. However, $w$ cannot be reduced much below the pump wavelength $w \sim 2\pi c/n\omega_{p_o}$ without causing a significant pump spillage into the superconductor. Figure 1(b) indicates that the region of highest efficiency for the single-step converter strongly overlaps with the region of highest thermal noise.

*Two-step quantum transduction.* The above analysis suggests that the noise photon population scales exponentially with the ratio $\omega_\mu/\omega_{p_o}$. Thus, the noise can be significantly reduced by introducing an intermediate frequency mode $s_i$ with a frequency $\omega_i$ such that $\omega_\mu \ll \omega_i \ll \omega_o$, and performing the transduction in two steps ($s_\mu$ to $s_i$, then $s_i$ to $s_o$, Fig. 2(a) and 2(b)). The fundamental advantage of this scheme comes from its intrinsic resources for minimizing thermal noise. Specifically, the first step ($s_\mu \rightarrow s_i$) is pumped by photons of relatively low energy $\hbar \omega_i \ll \hbar \omega_o$, which minimizes the associated heating. The second step ($s_i \rightarrow s_o$) starts from a frequency $\omega_i \gg \omega_\mu$, making it less sensitive to the heating. In addition, the second step's coupling strength scales proportionately to $\omega_i$, reducing the pump intensity requirement compared to the single-step conversion.

Quantum-limited frequency mixing between microwave and the mm-wave to THz range based on the superconductor-insulator-superconductor (SIS) mixers [31-36] and kinetic inductance (KI) nonlinearity has been extensively studied [25,26,37-39]. Here, we consider a KI-based mixer to implement the first conversion step ($s_\mu$ to $s_i$,) and an EO mixer as described earlier to implement the second step ($s_i$ to $s_o$), as shown in Fig. 2(a) and 2(b). Similarly, the intermediate pump signal $p_i$ (at $\omega_{p_i} = 2\pi \times (600 - 8)/2$ GHz) and $s_\mu$ interact to generate the intermediate state signal $s_i$. The cooperativity of the KI-mediated interaction is defined as $C_{KI} = 4g_{KI}^2|p_i|^4/\kappa_\mu\kappa_i$, where $\kappa_\mu$ and $\kappa_i$ are the loss rates of microwave mode and intermediate mode, respectively. The KI coupling rate $g_{KI}$ is defined as in Supplemental Material, Chapter IV. We analyze the performance of two cascaded transducer under the boundary condition $C_{KI} = C_{EO} = 1$, where we redefine $C_{EO} = 4g_{EO}^2|p_o|^2/\kappa_i\kappa_o$.

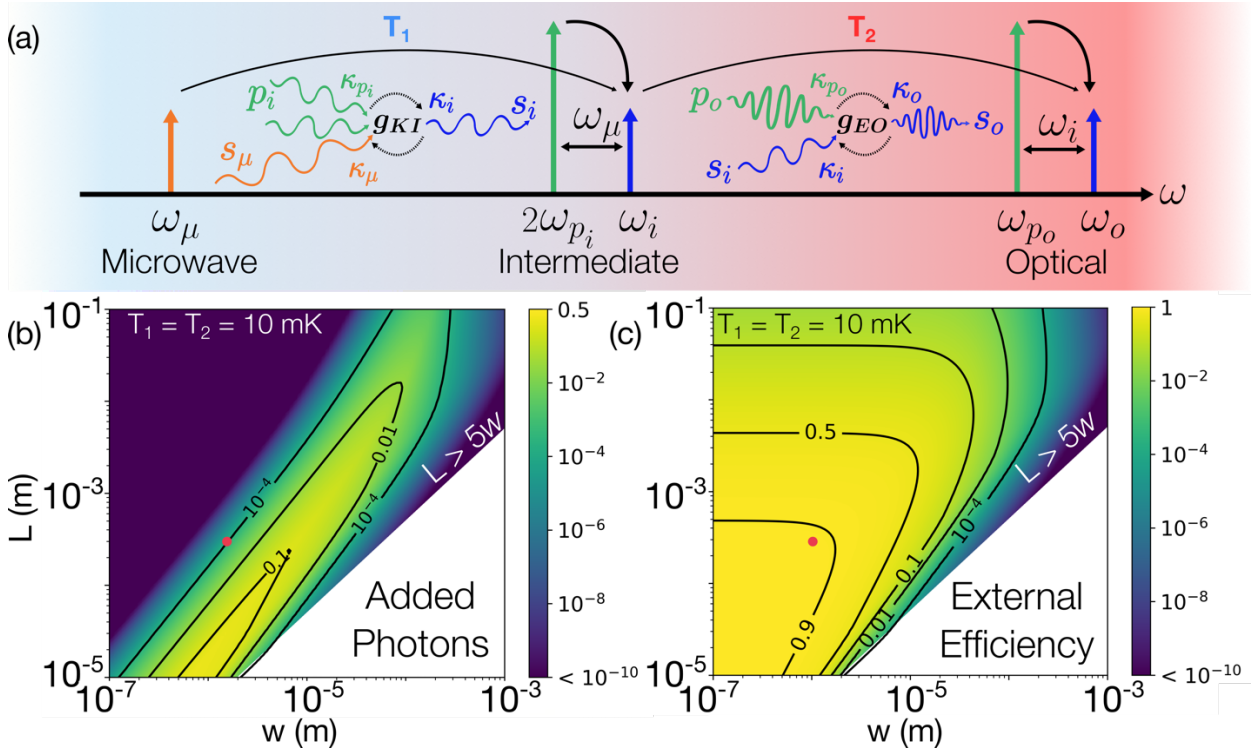

FIG. 2. (a) Description of two-step transduction using the kinetic inductance and electro-optic nonlinearities. Generalized heating model results for (b) the intermediate mode occupancy and (c) the external conversion efficiency for a device operating at $\omega_i = 2\pi \times 600$ GHz.

In Fig. 2(c) and 2(d) we plot the efficiency $\eta_2$ and thermal mode occupancy $n_{\mu,2}$ of the two-step converter with $\omega_i = 600$ GHz, as a function of the length $L$ and the width $w$ of the second-step converter. The estimated temperature increase in the KI step minimally affects the noise and efficiency of the scheme compared to the EO step, as demonstrated in Supplemental Material, Chapter VI. While the two steps may be placed separately at two different cryostat base temperatures ($T_1$ and $T_2$), for the purpose of comparing to a single-step transducer, we first consider the two based temperatures to be equal: $T_1 = T_2 = 10$ mK. We observe a multi-order reduction in noise photon occupancy $n_{\mu,2}$ compared to the single-step result $n_{\mu,1}$. Higher interaction strength in this region coincidentally reduces the optical pump power required for unit cooperativity, which allows the efficiency $\eta_2$ for a two-step transducer to be higher than $\eta_1$. In particular, the model predicts a quantum-limited performance with $\eta_2 \sim 93\%$ and $n_{\mu,2} \sim 10^{-8}$ at the point marked by the red dot, corresponding to $w = 1$ μm and $L = 300$ μm.

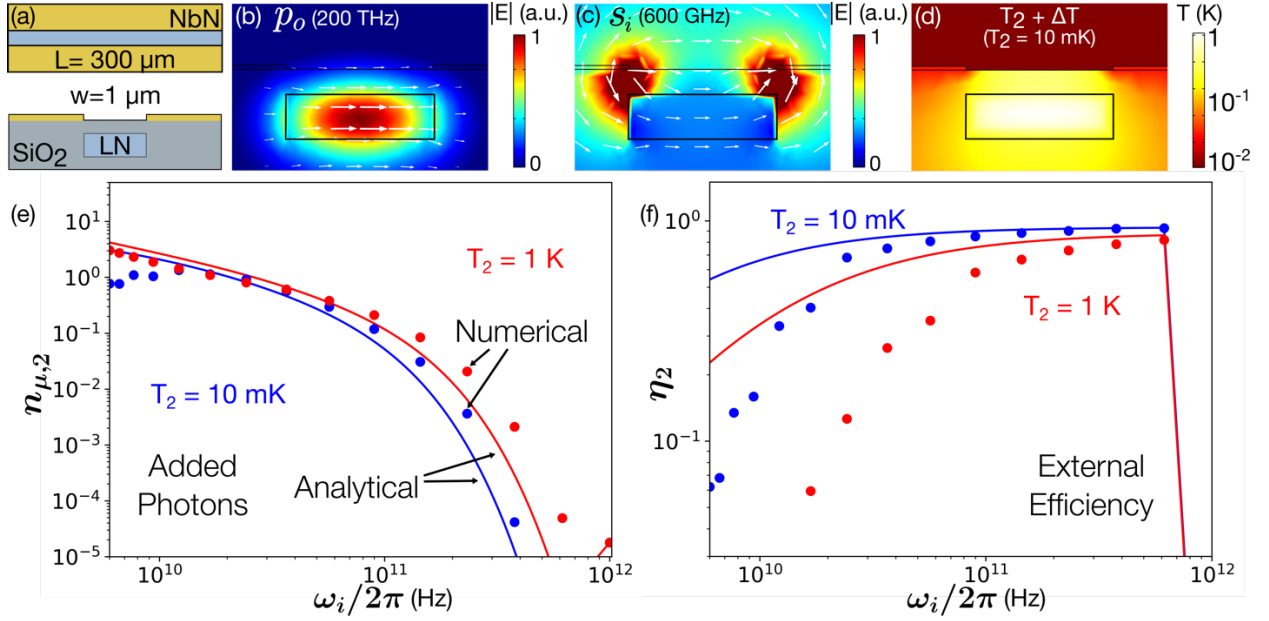

FIG. 3. (a) Top view and cross section of simulated LN structure integrated with superconducting NbN and field profiles for (b) $p_o$, (c) $s_i$, and (c) $\Delta T$. Colors in (b) and (c) represent the electric field norm in linear scale, and the arrows represent electric field strength and direction. Colors in (d) represent the temperature distribution in logarithmic scale. Analytical and numerical calculations of (e) mode occupancy and (f) external conversion efficiency of the EO step at 10 mK and 1 K with $w = 1$ µm and $L = 300$ µm.

To validate the analytical results, we performed finite element simulations of nonlinear wave dynamics and heat transfer in the EO step as a function of the intermediate frequency $\omega_i$ using COMSOL Multiphysics. The simulated structure consists of an LN waveguide buried in an SiO$_2$ cladding and embedded into a superconducting NbN coplanar waveguide (similar to structures in [22,23,40,41]) with dimensions $w = 1$ µm and $L = 300$ µm. Figures 3(a) and 3(b) show both the analytical and numerical estimations for $n_{\mu,2}$ and $\eta_2$, respectively. The optical waveguide is placed 0.5 µm below the superconducting film to minimize the absorption of optical photons by the superconductor. To phase-match the sum frequency generation between the modes $s_i$, $p_o$ and $s_o$, the LN waveguide thickness was adjusted to 300 nm [40,41]. The numerical model takes into account the different thermal conductivities for LN, SiO$_2$, air and NbN (see Supplemental Material, Chapter III).

Numerical results in Fig. 3(a) and 3(b) (blue dots) confirm the analytical trends described by equation 1 (blue solid lines). At higher intermediate frequencies $\omega_i > 2\pi \times 100$ GHz, the heating is strongly alleviated, resulting in near-unity external efficiency, and suppressing the noise $n_{\mu,2}$ by several orders of magnitude. The strong departure of the numerical data from the analytical trend at low $\omega_i$ in Figure 3(b) can be explained by the imperfect mode overlap between modes $s_i$ and $p_o$. At intermediate frequencies $\omega_i$ in the microwave range, the mode overlap is of the order 0.01, requiring a significantly stronger pump power compared to the analytical model. However, for $\omega_i \sim 1\ THz$ the mode overlap reaches 0.25, and the numerical results approach the predictions of the analytical model.

Since, $\omega_i$ can be chosen to be much higher than $\omega_\mu$, one can operate the EO transduction step at higher base temperatures than the KI step. Thus, the EO step can be placed away from the mK cryostat chamber, isolating the latter from all optical signals, and suppressing the effects of optical absorption such as quasiparticle scattering and backaction [42]. To verify this hypothesis, we performed numerical and analytical simulations of $\eta_2$ and $n_{\mu,2}$ at T$_2$ = 1 K (Figure 3 (a) and 3(b), red dots and red solid lines, respectively). We find that the efficiency and noise are still nearly quantum-limited at $\omega_i > 300$ GHz, with $\eta_2 > 90\%$ and $n_{\mu,2} < 10^{-4}$.

*Discussion.* Modular quantum architectures require coherently transferring quantum states between qubits and generating entanglement with rates higher than the inverse qubit coherence times [1]. Thanks to their unique intrinsic bandwidth in the MHz range and thermal performance, two-step electro-optic transducers can go beyond these rates. The performance of two-step electro-optic transducers can be improved using superconducting materials with critical temperatures higher than that of NbN. In particular, these materials may help further reduce losses at the intermediate frequency, and allow higher base temperatures, associated with more efficient cooling [43]. While we have only analyzed the scaling of thermal noise, spontaneous scattering is another commonly known noise source in quantum transducers. Spontaneous noise may be suppressed by using a resonant configuration [4,18-26] and/or by engineering the phase-matching through waveguide dispersion [44-46]. A more detailed analysis will be useful to extend these techniques to the two-step electro-optic transducers. Beyond quantum transduction protocols, low-noise superconductor-based electro-optic interfaces can help close the THz gap with applications in high-rate communications [47], sensing [48], and astronomy [49,50].

We acknowledge Sary Bseiso for his help in preparing the manuscript. F.S. and S.I.B. acknowledge the startup funding at the University of Illinois at Urbana-Champaign.

# Supplementary Materials for

# Terahertz-Mediated Microwave-to-Optical Transduction

Furkan Sahbaz[1,3], James N. Eckstein[2,3], Dale J. Van Harlingen[2,3] and Simeon I. Bogdanov[1,3]

[1]Department of Electrical and Computer Engineering, and Nick Holonyak, Jr. Micro and Nanotechnology Laboratory, University of Illinois at Urbana-Champaign, Urbana, Illinois 60801, USA

[2]Department of Physics and Materials Research Laboratory, University of Illinois at Urbana Champaign, Urbana, Ilinois 81801

[3]Illinois Quantum Information Science and Technology Center, University of Illinois Urbana-Champaign, Urbana, Illinois 61801, USA

## I. Electro-optic interaction between microwave and optical signals

The coupling between microwave, optical pump, and the optical sideband (as demonstrated in Fig. 1 of the main text) is described by the nonlinear coupling rate [S1]

$$g_{EO} = \sqrt{\frac{\hbar \omega_\mu \omega_{p_o} \omega_o}{8\varepsilon_0 \varepsilon_\mu \varepsilon_{p_o} \varepsilon_o}} \chi^{(2)} \frac{\xi}{\sqrt{V_\mu^{1/3} V_{p_o}^{1/3} V_o^{1/3}}} \quad (1)$$

where frequency $\omega_\mu$, $\omega_{p_o}$, $\omega_o$, relative permittivity $\varepsilon_\mu$, $\varepsilon_{p_o}$, $\varepsilon_o$ and effective mode volume $V_\mu$, $V_{p_o}$, $V_o$ are defined for each mode, i.e. microwave, optical pump and optical sideband. The overlap factor $\xi$ defined over the nonlinear region ($\chi^{(2)}$), and effective mode volumes are given by:

$$\xi = \frac{\int_{\chi^{(2)}} dV \psi_o^* \psi_\mu \psi_{p_o}}{\prod_{i=\{\mu,p_o,o\}} \left| \int_{\chi^{(2)}} dV \psi_i^* \psi_i \psi_i \right|^{1/3}} \quad (2)$$

$$V_i = \frac{\left( \int dV \varepsilon_i |\psi_i|^2 \right)^3}{\left| \int_{\chi^{(2)}} dV \varepsilon_i^{3/2} \psi_i^* \psi_i \psi_i \right|^2}, i = \{\mu, p_o, o\} \quad (3)$$

The cooperativity is defined as

$$C_{EO} = \frac{4g^2 n_p^2}{\kappa_\mu \kappa_o} = \frac{\hbar \omega_p}{2\varepsilon_0 \varepsilon_p V_{p_o}^{1/3}} \chi^{(2)2} \frac{\xi}{\varepsilon_\mu V_\mu^{1/3} \kappa_\mu} \frac{\xi}{\varepsilon_S V_o^{1/3} \kappa_o} \quad (4)$$

Where $\psi_\mu, \psi_{p_o}, \psi_o$ and $\kappa_\mu, \kappa_{p_o}, \kappa_o$ are electric field amplitudes and loss rates for the corresponding modes, respectively. The loss rate for each mode consists of the external loss rate and the internal loss rate, i.e. $\kappa = \kappa_{ex} + \kappa_{in}$.

## II. Defining the loss rates based on material- and geometry-dependent parameters

The internal loss rate of the superconducting material can be defined as

$$\kappa_{\mu,i} = \omega_\mu \frac{\sigma_1}{\sigma_2} \quad (5)$$

The complex conductivity ($\sigma = \sigma_1 - i\sigma_2$) of superconducting NbN is modeled from [S2-S7] by using material parameters and fitting relations. The external loss rate depends on the field radiation by the device, and therefore on the separation between superconducting films and the device length [S2,S8].

$$\kappa_{\mu,e} = \omega_\mu \left(\frac{w}{L}\right)^2 \quad (6)$$

$$\kappa_\mu = \omega_\mu \left(\frac{\sigma_1}{\sigma_2} + \left(\frac{w}{L}\right)^2\right) \quad (7)$$

where $L$ is the device length, and $w$ is the distance between superconducting films. Furthermore, the loss rate of the optical component can be calculated by eq. 8-10, where $c$ is the speed of light, $\alpha$ is absorption coefficient, $n_g$ is the optical group index and L corresponds the optical path [S9],

$$\kappa_{p_o,i} = \alpha c \quad (8)$$

$$\kappa_{p_o,e} = \frac{c}{n_g L} \quad (9)$$

$$\kappa_{p_o} = \alpha c + \frac{c}{n_g L} \quad (10)$$

### III. Optical and thermal material parameters

| Parameter | Value |
|---|---|
| Absorption coefficient ($\alpha$) of LiNbO$_3$ | 0.84 cm$^{-1}$ at 200 THz [S10]<br>2 cm$^{-1}$ - 5 cm$^{-1}$ up to 1.2 THz [S11] |
| Refractive index of LiNbO$_3$ | 2.3 at 200 THz [S10]<br>4.9 - 5 up to 1.2 THz [S11] |
| Nonlinear coefficient d$_{33}$ in LiNbO$_3$ | 27 pm · V$^{-1}$ [S12] |
| Thermal conductivity of SiO$_2$ | 0.0001 W·m$^{-1}$·K$^{-1}$ at 10 mK [S13]<br>0.01 W·m$^{-1}$·K$^{-1}$ at 1 K [S13] |
| Heat capacity of SiO$_2$ | 0.00001 J·kg$^{-1}$·K$^{-1}$ at 10 mK [S13]<br>0.0001 J·kg$^{-1}$·K$^{-1}$ at 1 K [S13] |
| Thermal conductivity of NbN | 0.005 W·m$^{-1}$·K$^{-1}$ at 10 mK [S14]<br>5 W·m$^{-1}$·K$^{-1}$ at 1 K [S14] |
| Heat capacity of NbN | $(0.0283 \cdot T + 0.0012 \cdot T^3)$ J·kg$^{-1}$·K$^{-1}$ [S15] |
| Thermal conductivity of LiNbO$_3$ | $4 \cdot T^3$ W·m$^{-1}$·K$^{-1}$ [S16] |
| Heat capacity of LiNbO$_3$ | $2.705 \cdot 10^{-4} \cdot T^3$ J·kg$^{-1}$·K$^{-1}$ [S16] |

TABLE 1. Material parameters used in analytical and numerical calculations.

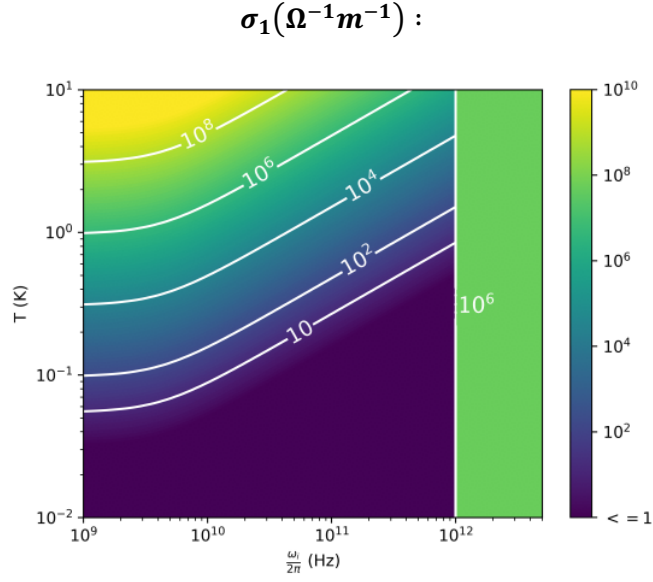

FIG I. Real part of conductivity from 1 mK to 10 K for NbN [S7].

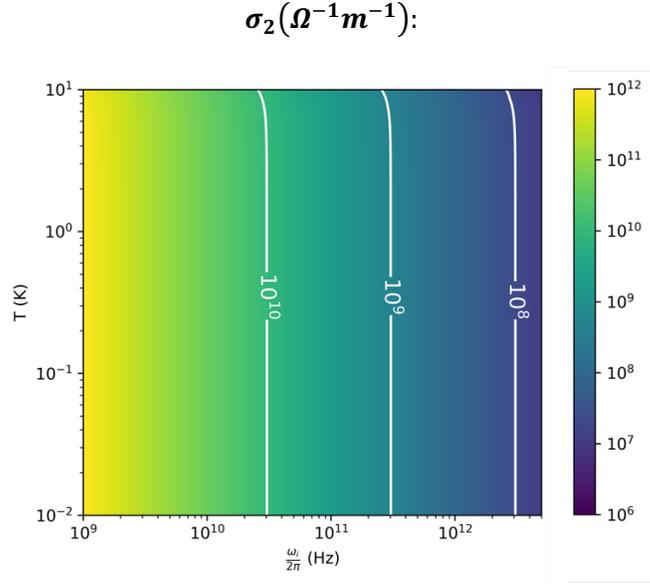

FIG II. Imaginary part of conductivity from 1 mK to 10 K for NbN [S7].

### IV. Defining kinetic inductance nonlinearity and coupling rate

The kinetic inductance (KI) based upconversion is a four-wave mixing process, where the pump mode is chosen such that $\omega_i = 2\omega_{p_i} + \omega_\mu$. The resulting cooperativity and coupling strength are given by [S17-S19]:

$$C_{KI} = \frac{4g_{KI}^2 n_p^2}{\kappa_\mu \kappa_i} \quad (11)$$

$$g_{KI} = \frac{3}{32} \frac{\hbar \omega_{p_i} \sqrt{\omega_\mu \omega_i}}{L_k I_*^2} \quad (12)$$

Here, $I_*$ and $L_k$ give the scaling current and kinetic inductance, respectively, which are defined by material- and geometry-dependent parameters as following:

$$I_* = \sqrt{\frac{\pi N_0 \Delta_0^3}{\hbar \rho}} wt \quad (13)$$

$$L_k = \frac{\hbar \rho}{\pi \Delta_0} \frac{L}{wt} \quad (14)$$

where $\Delta_0$, $N_0$ and $\rho$ are the gap energy, electron state density and normal resistivity of the superconducting materials. To maximize the KI nonlinearity, thickness of the films ($t$) is taken as 20 nm in agreement with [S18,S19].

### V. High-rate or continuous wave optical pumping temperature increase

In this section, we analyze the increase in nonlinear medium temperature caused by absorbing photons under continuous and pulsed schemes. Initially assuming continuously pumped photons incident on a volume with cross section $w^2$, length $L$, heat capacity $C_{th} = \rho c_{th} w^2 L$, and thermal conductance $G_{th} = g_{th} L$,

$$\Delta T_{in} = \left(\frac{E}{C_{th}}\right)(1 - e^{-\tau_{th}\kappa_{p_o,i}}) \quad (15)$$

gives the increase in medium temperature [S20,S21]. Here, $\tau_{th} = \frac{C_{th}}{G_{th}}$ is the thermal diffusion time constant and the optical photon absorption rate is $\kappa_{p_o,i}$. Energy of the absorbed photons is given by $E = |p_o|^2 \hbar \omega_{p_o}$, where $|p_o|^2$ is the number of incident optical photons. For a lithium niobate waveguide with $w = 1$ μm and $L = 300$ μm, the heating occurs on μs timescales. Therefore, assuming the heat is distributed radially through the volume ($L \gg w$), we can approximate the maximum temperature increase as:

$$\Delta T \sim \left(\frac{E \kappa_p \tau_{th}}{C_{th}}\right) = \frac{|p_o|^2 \hbar \omega_{p_o} \kappa_{p_o}}{g_{th} L} \quad (16)$$

### VI. Pump-induced heating for unity internal efficiency

For unit cooperativity we can impose a constraint for the number of photons that would allow the system for both EO and KI operations:

$$|p_o|^2 = \frac{\kappa_i \kappa_o}{4 g_{EO}^2} \quad (17)$$

$$|p_i|^2 = \sqrt{\frac{\kappa_\mu \kappa_i}{4 g_{KI}^2}} \quad (18)$$

Equations 1-10 provide scaling laws for loss and coupling rates. The pump photon-induced temperature increase under the condition of unit cooperativity is therefore given as:

$$\Delta T_{EO} = \left(\frac{\kappa_i \kappa_o}{4 g_{EO}^2} \cdot \frac{\hbar \omega_{p_o} \kappa_{p_o,i}}{g_{th} L}\right) \quad (19)$$

$$\Delta T_{KI} = \left( \sqrt{\frac{\kappa_\mu \kappa_i}{4 g_{KI}^2}} \cdot \frac{\hbar \omega_{p_i} \kappa_{p_i,i}}{g_{th} L} \right) \quad (20)$$

Therefore, by using eq. 19-20 and solving for the interaction volume, we can conclude the scaling of temperature increase for keeping 100% internal efficiency:

$$\Delta T_{EO} = \frac{2\varepsilon_0 \varepsilon_i \varepsilon_o \varepsilon_{op} \alpha c}{kT^n (\chi^{(2)} \xi)^2 \omega_{p_o}} \cdot \frac{V}{L} \cdot \left[ \frac{\sigma_{1,i}}{\sigma_{2,i}} + \frac{w^2}{L^2} \right] \cdot \left[ \alpha c + \frac{c}{nL} \right] \quad (21)$$

$$\Delta T_{KI} = \frac{N_0 \Delta_0^2 \omega_{p_i}}{3kT} \cdot \frac{V}{L} \cdot \frac{\sigma_{1,p}}{\sigma_{2,p}} \cdot \sqrt{\left( \frac{\sigma_{1,i}}{\sigma_{2,i}} + \frac{w^2}{L^2} \right) \cdot \left( \frac{\sigma_{1,\mu}}{\sigma_{2,\mu}} + \frac{w^2}{L^2} \right)} \quad (22)$$

$\Delta T_{EO}$ and $\Delta T_{KI}$ data for an intermediate frequency of $2\pi \times 600\ GHz$ are calculated below to demonstrate the significant difference between temperature increase caused by the kinetic inductance-based conversion and electro-optic conversion:

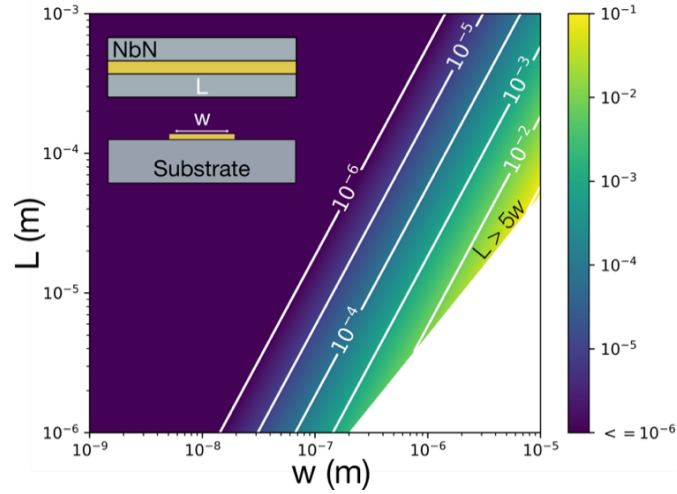

FIG III. $\Delta T_{KI}$ depending on $w$ and $L$. Inset is the kinetic inductor geometry with 20 nm thickness.

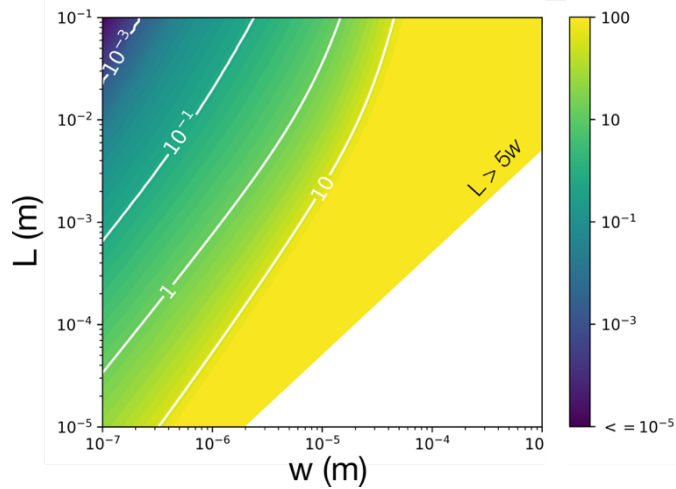

FIG IV. $\Delta T_{EO}$ depending on $w$ and $L$.